\documentclass[%
 aip,
 amsmath,amssymb,
 reprint,%
]{revtex4-2}

\usepackage{graphicx}% Include figure files
\usepackage{dcolumn}% Align table columns on decimal point
\usepackage{bm}% bold math
\usepackage[utf8]{inputenc}
\usepackage[T1]{fontenc}
\usepackage{mathptmx}
\usepackage{xcolor}

\begin{document}

\title{Room temperature Mott metal-insulator transition in V\textsubscript{2}O\textsubscript{3}
compounds induced via strain-engineering}

%\author{P. Homm$^1$, M. Menghini$^{1,2,*}$, J. W. Seo$^3$, S. Peters$^4$, J.-P. Locquet$^{1,*}$}
%\address{$^1$Department of Physics and Astronomy, KU Leuven, Celestijnenlaan 200D, 3001 Leuven, Belgium} 
%\address{$^2$IMDEA Nanociencia, Cantoblanco, E-28049, Madrid, Spain}
%\address{$^3$ Department of Materials Engineering, KU Leuven, Kasteelpark Arenberg 44, 3001 Leuven, Belgium}
%\address{$^4$SENTECH Instruments GmbH, Schwarzschildstr. 2,
%12489 Berlin, Germany}
%\email{mariela.menghini@kuleuven.be, jeanpierre.locquet@kuleuven.be}

\author{P. Homm}
 \affiliation{Department of Physics and Astronomy, KU Leuven, Celestijnenlaan 200D, 3001 Leuven, Belgium}
\author{M. Menghini}
% \altaffiliation[Also at ]{IMDEA Nanociencia, Cantoblanco, E-28049, Madrid, Spain}%
 \email{mariela.menghini@imdea.org.}
 \affiliation{Department of Physics and Astronomy, KU Leuven, Celestijnenlaan 200D, 3001 Leuven, Belgium}%
  \affiliation{IMDEA Nanociencia, Cantoblanco, E-28049, Madrid, Spain}
\author{J. W. Seo}
\affiliation{ Department of Materials Engineering, KU Leuven, Kasteelpark Arenberg 44, 3001 Leuven, Belgium}
\author{S. Peters}
\affiliation{ SENTECH Instruments GmbH, Schwarzschildstr. 2, 12489 Berlin, Germany}
\author{J.-P. Locquet} 
 \email{jeanpierre.locquet@kuleuven.be.}
 \affiliation{Department of Physics and Astronomy, KU Leuven, Celestijnenlaan 200D, 3001 Leuven, Belgium}

\date{\today}% It is always \today, today,

\keywords{metal-insulator transition, strain engineering, vanadium oxide}

\begin{abstract}

Vanadium sesquioxide (V\textsubscript{2}O\textsubscript{3})
is an archetypal Mott insulator in which the atomic positions and electron correlations change as temperature, pressure
or doping are varied giving rise to different structural, magnetic or electronic phase transitions. Remarkably, the isostructural Mott  transition in  Cr-doped V\textsubscript{2}O\textsubscript{3}
between  paramagnetic metallic and insulating phase observed in bulk has been elusive in thin film compounds so far.

Here, via continuous lattice deformations induced by heteroepitaxy we demonstrate
a room temperature Mott metal-insulator transition in 1.5\%
Cr-doped and pure V\textsubscript{2}O\textsubscript{3} thin films.
By means of a controlled epitaxial strain, not
only the structure but also the intrinsic electronic and optical properties
of the thin films are stabilized at different intermediate states
between the metallic and insulating phases, inaccessible in bulk materials. This leads
to films with unique features such as a colossal change in room temperature resistivity ($\Delta$R/R
up to 100,000 \%) and a broad range
of optical constant values, as consequence of a strain-modulated
bandgap. We propose a new phase diagram for pure and Cr-doped V\textsubscript{2}O\textsubscript{3} thin films with the engineered in-plane lattice constant as a  tuneable  parameter.
Our results demonstrate that controlling phase transitions in correlated systems by epitaxial
strain offers a radical new approach to create the next generation of Mott devices.

\end{abstract}

\maketitle
%\begin{onehalfspace}
\section{Introduction}

Vanadium sesquioxide (V\textsubscript{2}O\textsubscript{3}) is an
example of a Mott insulator presenting both non-isostructural and isostructural
metal\textendash insulator transitions (MIT) as a function of temperature,
doping and pressure \cite{McWhan1973,McWhan1970,McWhan1969} and has attracted great interest for applications.
V\textsubscript{2}O\textsubscript{3} undergoes a non-isostructural
MIT at 160 K upon cooling from a corundum paramagnetic metallic (PM)
phase to a monoclinic antiferromagnetic insulating phase (AFI) \cite{Frandsen2016,Dernier1970}, yielding a
resistivity change of 7 orders of magnitude \cite{McWhan1973}.
In contrast to this low-temperature MIT, the isostructural Mott MIT between paramagnetic
metallic (PM) and insulating (PI) phases occurs close to room temperature
(RT) and has been observed for a limited range of Cr doping (0.5\% \textendash{} 1.7\%) concentration \cite{Metcalf2007}.
This transition has been observed in bulk crystals with a resistivity change of 3 orders of magnitude
and a variation in \textit{a}-axis lattice parameter of 1\% within the same corundum
structure \cite{McWhan1973,McWhan1970}.

The bulk phase diagram of V\textsubscript{2}O\textsubscript{3}
as a function of temperature, doping and applied pressure has been well
established based on extensive experimental data from single crystals \cite{Lupi2010,Mansart2012,Rodolakis2010,Rodolakis2011},
and supported by a number of theoretical calculations that successfully
predicted the density of states (DOS) of the different phases \cite{Hansmann2013,Guo2014,Basov2011}.
In thin films, various deviations from the bulk phase diagram have been observed. For example, recently it has been shown that the AFI phase in thin films is characterized by stripe domains related to different lattice distortions \cite{Ronchi2019}. 
Then, the low-temperature  PM-AFI transition has been reported by many groups \cite{Metcalf2007}, while the isostructural RT Mott transition involving PM-PI phases - despite greater interest for applications - has been missing in most studies \cite{Metcalf2007,Querre2016,Homm2015}. This suppression of the RT MIT has been attributed to the clamping effect
of the substrate in epitaxial films \cite{Homm2015} or to structural disorder and small grain size present in polycrystalline films \cite{Metcalf2007,Querre2016}.
Especially, epitaxial strain has so far shown its strong effect on the phase transitions in the vanadate family, which can range from a change in transition temperature \cite{Aetukuri2013} or direction dependent electrical properties \cite{Liang2019} in VO\textsubscript{2} films to a complete suppression of the low-temperature phase transition \cite{Autier-Laurent2006} in V\textsubscript{2}O\textsubscript{3} films.

Salev et al. \cite{Salev2019} demonstrated that the low-temperature MIT can actively be modulated through a voltage-actuated PMN-PT
{[}Pb(Mg,Nb)O\textsubscript{3}-PbTiO\textsubscript{3}{]} ferroelectric actuator, which transfers switchable ferroelastic strain into the epitaxial V\textsubscript{2}O\textsubscript{3} film and leads to a giant nonvolatile resistive switching of $\Delta$R/R \textasciitilde{} 1,000\%, while the effect at room temperature is only $\Delta$R/R \textasciitilde{} 18\%. 
Querr$\acute{e}$ et al. \cite{Querre2018} observed non-volatile resistive switching in Cr-doped V\textsubscript{2}O\textsubscript{3} films, where conducting filaments form as a result of a local electric field driven low-temperature Mott transition. These recent achievements have pointed out the great potential of Mott insulator thin films with large resistivity changes across phase transitions for application in sensors and memory devices.
However, triggering the metal-insulator phase transition at room temperature would be even more desirable for their implementation in devices.

The RT MIT has been observed in polycrystalline thin films of Cr-doped V\textsubscript{2}O\textsubscript{3} by optical transmission measurements while electrical resistivity didn't show a transition \cite{Metcalf2007}. Epitaxial V\textsubscript{2}O\textsubscript{3} thin films showed a local, pressure-induced, transition at RT  by means of a contact AFM tip \cite{Alyabyeva2018} whereas a systematic study by varying the oxygen doping concentration in epitaxial films didn't show the RT MIT \cite{thorsteinsson2018}. These results indicate that the RT MIT can be stabilized within a narrow window of conditions, which need to be systematically explored.
So far, a generalized phase diagram for V\textsubscript{2}O\textsubscript{3} thin films does not exist but would be highly desirable to gain fundamental understanding about the interplay between epitaxial strain and Mott transition and to use these films in practical applications.

Here, we present a systematic strain-engineering study of  1.5\% Cr-doped and pure V\textsubscript{2}O\textsubscript{3} thin films.
Our approach is based on the use of precisely engineered oxide heterostructures as artificial substrates to systematically tune the in-plane lattice parameter in the pure and 1.5\% Cr-doped V\textsubscript{2}O\textsubscript{3} thin films within and beyond the reported structural states in the bulk.
The in-plane lattice parameters of the engineered substrates were gradually adjusted between 4.943 {\AA} and 5.037 {\AA} by varying the precise composition of the thin film buffer layer.
Detailed chemical analysis by means of Energy Dispersive X-ray spectroscopy (EDX) confirmed the chemical composition of the buffer layers and the V\textsubscript{2}O\textsubscript{3} films. Extensive structural studies including High Resolution X-Ray Diffraction (HRXRD), Reciprocal Space Maps (RSMs) and High Resolution Transmission Electron Microscopy (HRTEM) revealed the in-plane lattice parameters of the heterostructures and the amount of induced stress. 
Temperature-dependent electrical resistivity curves were measured using the Van Der Pauw (VDP) configuration. In addition, optical properties
were addressed by Spectroscopic Ellipsometry (SE) to illustrate how physical properties in the V\textsubscript{2}O\textsubscript{3} thin films can be fully controlled by epitaxial strain.

The results show an effective transfer of strain achieved by  heteroepitaxy leading to a Mott MIT in a wide range of temperatures around RT in 1.5\% Cr-doped and pure V\textsubscript{2}O\textsubscript{3}
compounds.
We observed a colossal change in room temperature resistivity (RTR) with $\Delta$R/R up to \textasciitilde{} 100,000\% as well as a significant change in optical constants associated with the strain-modulated bandgap of these Mott insulators. Based on these results, a novel phase diagram of V\textsubscript{2}O\textsubscript{3} thin films was proposed with the RT in-plane parameter as variable.
These results demonstrate that meticulous strain engineering is a powerful tool to tune physical properties of strongly-correlated electron systems and also allows accessing novel states unaccesible in bulk. 
The stabilization and control of the MIT at RT in 1.5\% Cr-doped and pure V\textsubscript{2}O\textsubscript{3} thin films is an important achievement towards implementation of Mott materials in devices. 

\section{Experimental}

\subsection{Thin film growth} 
All the thin film layers were deposited on (0001)-Al\textsubscript{2}O\textsubscript{3}
substrates by oxygen-assisted molecular beam epitaxy (MBE). The vacuum chamber
had a base pressure of 10 \textsuperscript{\textminus 9}
Torr and it was equipped with RHEED to confirm the epitaxial growth
\textit{in situ}. The oxide heterostructures
consist of a 40 nm Cr\textsubscript{2}{O}\textsubscript{3}
epitaxial layer followed by a 30 nm thick (Cr\textsubscript{$1-y$}{\footnotesize{}Fe}\textsubscript{$y$}{\footnotesize{})}\textsubscript{2}{O}\textsubscript{3}
solid solution (y = 0 corresponds to a 40 nm Cr\textsubscript{2}{O}\textsubscript{3}
only). To guarantee that all V\textsubscript{2}{O}\textsubscript{3}
film compounds have the same chemical interface with the heterostructure,
an ultrathin 3 nm Cr\textsubscript{2}{O}\textsubscript{3}
layer was deposited on the (Cr\textsubscript{$1-y$}{\footnotesize{}Fe}\textsubscript{$y$}{\footnotesize{})}\textsubscript{2}{O}\textsubscript{3}
alloys with y > 0. Finally, the 60 nm 1.5\% Cr-doped and pure V\textsubscript{2}{O}\textsubscript{3}
films were deposited on the artificial substrates composed by the
oxide heterostructures.

\subsection{Structural characterization}
The structural properties of the samples were characterized by means of high-resolution x-ray diffraction (XRD)
with a Panalytical X\textquoteright pert Pro diffractometer using
monochromatic K\textsubscript{{\footnotesize{}$\alpha$1}}
radiation of a Cu anode. Out-of-plane measurements in the $\theta$/2$\theta$
configuration were done around the symmetric peak ($0$$0$$0$$6$)
and 2-axes measurements - reciprocal space maps (RSM) - around the
asymmetric peak ($1$ $0$ $\bar{1}$ $\underline{10}$)
were performed to extract in-plane and out-of-plane lattice spacings.
Besides, the structural quality and composition were also addressed by transmission electron microscopy (TEM) and Energy dispersive
X-ray spectroscopy (EDX), respectively. TEM and EDX analysis were
performed using a probe-lens corrected JEOL ARM200F scanning transmission
electron microscope operating at 200 kV and equipped with a cold-field
emission gun and a Centurio energy dispersive x-ray spectrometer.
The TEM samples were prepared by means of a focused ion beam (FIB).

\subsection{Electrical and optical characterization}
Temperature-dependent resistivity measurements in the Van der Pauw
(VDP) configuration were performed in an Oxford Optistat CF2-V cryostat
using a Keithley 4200-SCS parameter analyzer. The temperature sweep
rate was of 1 K per minute. Finally, spectroscopic ellipsometry (SE) measurements from 70 meV to 0.5 eV in the MIR and from 0.5 eV (NIR)
to 6 eV (UV) were performed at SENTECH Instruments GmbH in Berlin
with a SENDIRA and a SENresearch 4.0 SER 850 DUV ellipsometers, respectively.
The ellipsometer angles $\psi$ and $\Delta$
were measured at room temperature using three different angles of
incidence ($65^{\circ}$, $70^{\circ}$  and $75^{\circ}$ ). Independent measurements on the layers
in the oxide heterostructures were performed before the 1.5\% Cr-doped
and pure V{\scriptsize{}}\textsubscript{2}{O}{\scriptsize{}}\textsubscript{3}
films were deposited and the data were fitted applying the Tauc-Lorentz
oscillator model \cite{Jellison1996}. Then, the measurements were
repeated on the stack with the top 1.5\% Cr-doped or pure V\textsubscript{2}{O}\textsubscript{3}
films and the data were fitted applying the same optical model used for analyzing the layers in the heterostructure.

\section{Results and discussion}

% Figure Structural Properties ---------------

\begin{figure*} [hhh]
\includegraphics[scale=0.3]{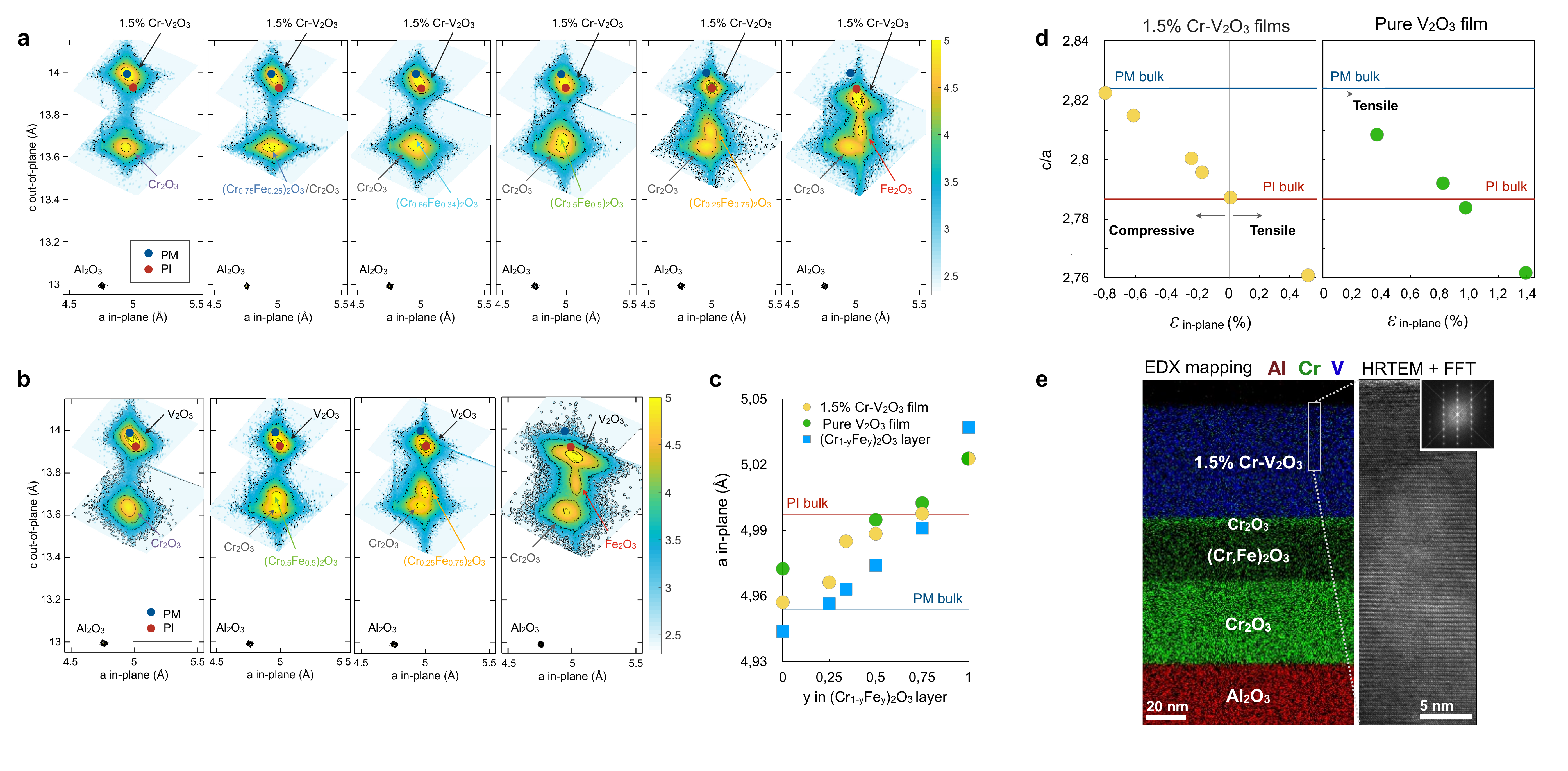}
\caption{Structural properties of strained 1.5\%
Cr-doped and pure V\protect\textsubscript{2}O\protect\textsubscript{3}
films. \textbf{a, b} RSMs around the ($1$ $0$ $\bar{1}$ $\underline{10}$)
reflections transformed to in-plane and out-of-plane lattice spacings
of the 1.5\% Cr-doped and pure V\protect\textsubscript{2}O\protect\textsubscript{3}
films grown on the different (Cr\protect\textsubscript{$1-y$}Fe\protect\textsubscript{$y$})\protect\textsubscript{2}O\protect\textsubscript{3}/Cr\protect\textsubscript{2}O\protect\textsubscript{3}
heterostructures. The PM and PI bulk lattice parameters are indicated
by blue and red dots, respectively. \textbf{c} In-plane lattice parameters
of the (Cr\protect\textsubscript{$1-y$}Fe\protect\textsubscript{$y$})\protect\textsubscript{2}O\protect\textsubscript{3}
alloy layers in the oxide heterostructures (ligh blue squares) and
the corresponding ones for the 1.5\% Cr-doped (yellow circles) and
pure (green circles) V\protect\textsubscript{2}O\protect\textsubscript{3}
films. \textbf{d} c/a ratio as a function of the in-plane strain ($\epsilon$\protect\textsubscript{in-plane})
induced by heteroepitaxy onto the 1.5\% Cr-doped and pure V\protect\textsubscript{2}O\protect\textsubscript{3}
films. In \textbf{c} and \textbf{d}, PM and PI bulk values are indicated
by the blue and red lines, respectively. \textbf{e} (Left) Color-coded
EDX mapping of a stack representative of the 1.5 \% Cr-doped V\protect\textsubscript{2}O\protect\textsubscript{3}
film series and (Right) HRTEM image of the top layer and the corresponding
Fast-Fourier Transform (FFT). }
\label{Structural figure}
\end{figure*}

%  ---------------

%\subsection{Oxide heterostructure and final stacks.} 

Systematic straining of 1.5\% Cr-doped and pure V\textsubscript{2}O\textsubscript{3}
films was achieved by using oxide heterostructures consisting of several epitaxial thin films with corundum structure. The thickness of the 1.5 \% Cr-doped and pure V\textsubscript{2}O\textsubscript{3} layers was kept to 60 nm.
As there is a large lattice mismatch of > 4\% between Al\textsubscript{2}O\textsubscript{3}
and Cr-doped and pure V\textsubscript{2}O\textsubscript{3} \cite{Homm2015},
a 40 nm buffer layer of Cr\textsubscript{2}O\textsubscript{3} was
first deposited. As demonstrated previously \cite{Dillemans2014}, Cr\textsubscript{2}O\textsubscript{3} buffer layer can grow almost relaxed on 
Al\textsubscript{2}O\textsubscript{3} and enables the subsequent  growth of V\textsubscript{2}O\textsubscript{3} structurally decoupled from the substrate
with a lattice mismatch of about 0.1\%.

In addition, Cr\textsubscript{2}O\textsubscript{3} has the additional advantage that it can be alloyed with Fe.
(Cr\textsubscript{$1-y$}Fe\textsubscript{$y$})\textsubscript{2}O\textsubscript{3} also forms a corundum structure with the in-plane lattice parameter that can be changed by more than 1\% (between \textit{a} = 4.959 \AA$\,$   and \textit{a} = 5.035 \AA) \cite{Mashiko2011} by
varying the Fe content y. On top of this (Cr\textsubscript{$1-y$}Fe\textsubscript{$y$})\textsubscript{2}O\textsubscript{3} layer,
an ultrathin Cr\textsubscript{2}O\textsubscript{3} layer (3 nm) was deposited that remained
coherent/pseudomorphic with the underlying (Cr\textsubscript{$1-y$}Fe\textsubscript{$y$})\textsubscript{2}O\textsubscript{3}
layer. This ultrathin Cr\textsubscript{2}O\textsubscript{3} layer ensured that the 1.5\% Cr-doped and pure V\textsubscript{2}O\textsubscript{3 } films deposited on the buffer stack experienced exactly the same chemical interface so that possible chemical effects of the underlying layer can be excluded.
Figure \ref{Structural figure}\textbf{e} shows a cross-section EDX map obtained from a 1.5\% Cr-doped V\textsubscript{2}O\textsubscript{3 } film grown on the buffer stack described above. The color code linked to different Cr concentration in the layers reflects the layout of the final heterostructure used for this study. 
The Cr doping of 1.5\% was specifically chosen to ensure the presence of a single PI phase at RT while
in pure V\textsubscript{2}O\textsubscript{3} a single PM phase is expected \cite{McWhan1973,McWhan1970}. This allows us to address the
effect of strain engineering on both phases involved in the Mott transition.

The in-plane and out-of-plane lattice parameters of the strained (doped and undoped) V\textsubscript{2}O\textsubscript{3} films were determined by means of extensive HRXRD analysis.
Figures \ref{Structural figure}\textbf{a} and \textbf{b} show the RSMs of the 1.5\% Cr-doped and pure V\textsubscript{2}O\textsubscript{3 } films, measured around the ($1$ $0$ $\bar{1}$ $\underline{10}$) peak of each layer in the stacks. 
They clearly demonstrate that all layers in the stacks are single-phase, epitaxial and preserve the corundum structure. No secondary phases were detected.
The excellent structural quality is evidenced by the presence
of finite-size oscillations in out-of-plane $\theta$/2$\theta$
XRD measurements (see Fig. S1 in the supplementary information).

The ($1$ $0$ $\bar{1}$ $\underline{10}$) peak positions of the Cr\textsubscript{2}O\textsubscript{3}/(Cr\textsubscript{$1-y$}Fe\textsubscript{$y$})\textsubscript{2}O\textsubscript{3}/Cr\textsubscript{2}O\textsubscript{3} stack - and the lattice parameters (\textit{a}
and \textit{c}) - confirm that the engineered buffer has an in-plane lattice parameter effectively decoupled from the substrate.
The blue and red dots in the RSMs in Fig. \ref{Structural figure}\textbf{a} and \textbf{b} mark the bulk lattice parameters of the PM and PI phases, respectively.
By increasing the Fe content in the (Cr\textsubscript{$1-y$}Fe\textsubscript{$y$})\textsubscript{2}O\textsubscript{3}, the lattice parameters \textit{a} and \textit{c} of the 1.5\% Cr-doped and pure V\textsubscript{2}O\textsubscript{3} films gradually move from the blue dot position (corresponding PM with \emph{a}\textsubscript{\emph{PM}}\emph{, c}\textsubscript{\emph{PM}}) to the red dot position (corresponding PI with \emph{a}\textsubscript{\emph{PI}}\emph{, c}\textsubscript{\emph{PI}}). For the Fe concentration of $y=1$ (Fe\textsubscript{2}O\textsubscript{3}), the lattice parameters go even beyond the PI bulk value.

In-plane lattice parameters of the films are calculated based on the RSM data and summarised in Fig. \ref{Structural figure}\textbf{c} as a function of the Fe content $y$ together with the corresponding values for the buffer layers. The data show that the 1.5\% Cr-doped and pure V\textsubscript{2}O\textsubscript{3} films roughly follow the in-plane lattice parameter evolution of the buffer layers. It is important to note that no systematic widening of the RSM peaks is observed  as the Fe content  changes (except for the cases with $y =1$) in the buffer layers. Moreover, by using the Scherrer equation for the out-of-plane diffraction peaks we found that the crystalline coherence is almost the same as  the thickness extracted from x-ray reflectivity measurements (only an average difference of 3\%) evidencing a good uniformity of the films lattice structure. These results indicate that  the designed buffer layers are sufficient to grow films with a well-defined lattice constant across the whole film. In the case of $y =1$ the observed RSM peak  broadening could be due to local inhomogeneities or oxygen vacancies that can be more favorable due to the larger tensile strain in these films.

The in-plane lattice parameter of the (Cr\textsubscript{$1-y$}Fe\textsubscript{$y$})\textsubscript{2}O\textsubscript{3} layer (light-blue squares)
increases linearly with y (only the data point for $y=1$ deviates from this linear trend), evidencing that the produced buffer effectively tailors the amount of induced strain as a function of Fe content.
From the in-plane lattice parameters, the resulting in-plane
strain was calculated as  $\epsilon$\textsubscript{in-plane}=
(\textit{a}\textsubscript{\textit{film}} - \textit{a}\textsubscript{\textit{bulk}})/\textit{a}\textsubscript{\textit{bulk}},
with \textit{a}\textsubscript{\textit{bulk}} equal to
\textit{a}\textsubscript{\textit{PI}} (\emph{a}\textsubscript{\emph{PM}}) for  1.5\% Cr-doped 
(pure) V\textsubscript{2}O\textsubscript{3} films. Figure \ref{Structural figure}\textbf{d} shows the evolution of the  \textit{c/a} ratio as a function of the strain in the Cr-doped (left) and pure (right) V\textsubscript{2}O\textsubscript{3} films.
The continuous decrease of  \textit{c/a} between the PM and PI bulk values (marked by blue and red line, respectively) with $\epsilon$\textsubscript{in-plane} indicates that the \textit{c}-axis adjusts to the imposed in-plane deformations. The measured changes in the \textit{c/a}  ratio of our strained films slightly deviate from what is expected considering only an elastic deformation: The Poisson ratio for the PI phase in 1.5\% Cr-V\textsubscript{2}O\textsubscript{3} for $\epsilon$\textsubscript{in-plane} = -0.8 yiels \textit{c/a} = 2.831 \cite{Yang1985} while we find \textit{c/a} = 2.822 experimentally. The most remarkable result is the continuous change in \textit{c/a} between the PM and PI phases, in contrast to the discontinuous  one observed in single crystals \cite{McWhan1970}. This deviation indicates that the bulk and thin films of doped and pure V\textsubscript{2}O\textsubscript{3} behave differently allowing us to explore states with intermediate lattice parameters  that are not stable in bulk samples. As highlighted in Fig. \ref{Structural figure}\textbf{d}, in pure V\textsubscript{2}O\textsubscript{3} films only tensile
strain is possible because the Cr\textsubscript{2}O\textsubscript{3} layer has nearly the
same in-plane lattice parameter as the PM phase, and alloying with Fe leads
 to  larger in-plane parameters.
On the contrary, both compressive and tensile strains are accessible for the 1.5\%
Cr-doped V\textsubscript{2}O\textsubscript{3} films, which should be in the PI phase in the unstrained bulk phase.

Figure \ref{Structural figure}\textbf{e} shows a representative color-coded EDX map of the entire stack illustrating the sharp
interfaces present across the different layers in the oxide heterostructure. Comparable results were obtained for 1.5\% Cr-doped
(pure) V\textsubscript{2}O\textsubscript{3} films on the different buffers. 
In particular, systematic EDX line-scans across the interface between the ultrathin 3 nm Cr\textsubscript{2}O\textsubscript{3}
layer and the 1.5\% Cr-doped and pure V\textsubscript{2}O\textsubscript{3} layers confirmed that there is no Cr diffusion into the top vanadium oxide layer.
The Cr content in the Cr-doped films was estimated to be 1.5\% \textpm{} 0.3\%. Also the HRTEM image and corresponding Fast Fourier Transform (FFT) (shown in the inset)  demonstrate the single-crystalline character of the top layer across its whole thickness. The lattice spacings determined by HRTEM were in agreement with those extracted from HRXRD (Fig. \ref{Structural figure}\textbf{c}).

\begin{figure*}
\includegraphics[scale=0.28]{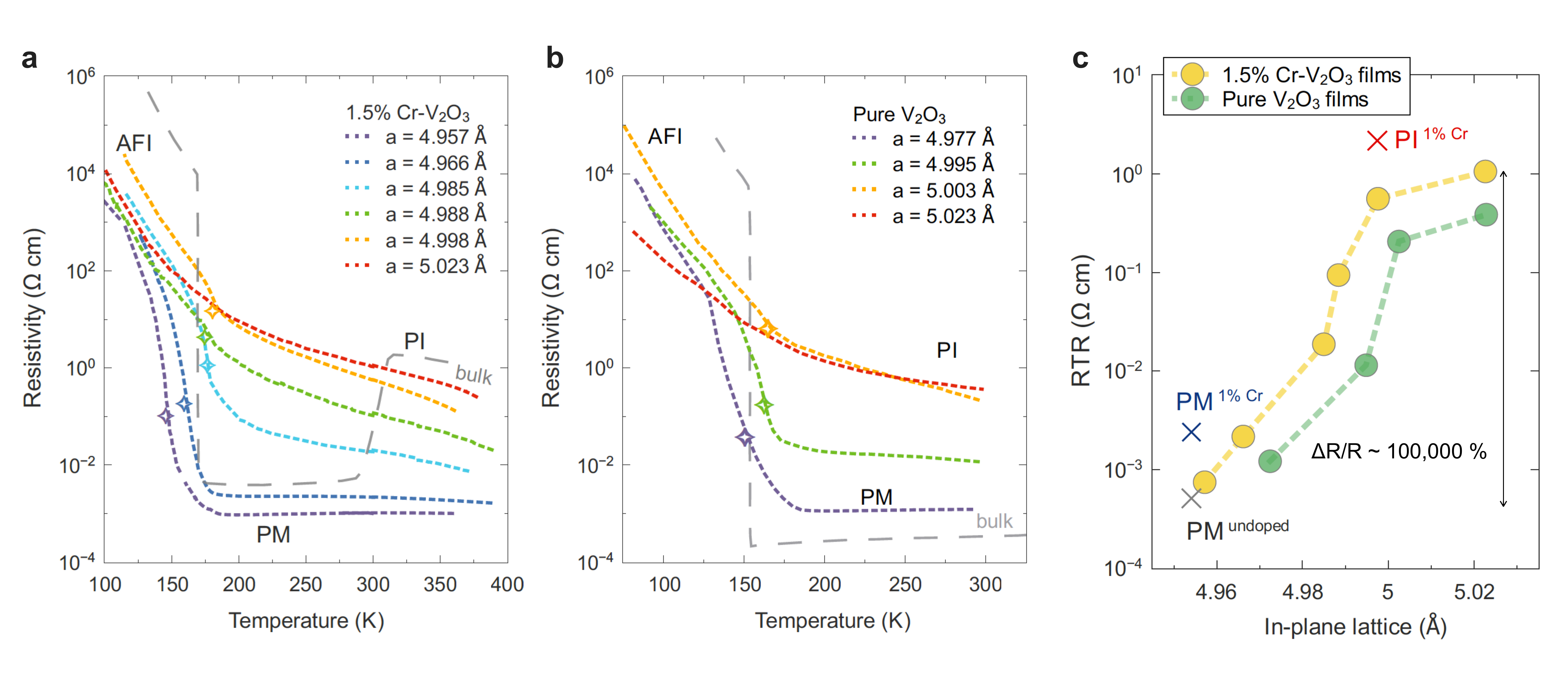}
\caption{\label{Electrical figure}Electrical properties of strained 1.5\%
Cr-doped and pure V\protect\textsubscript{2}O\protect\textsubscript{3}
films. \textbf{a, b} Resistivity curves R(T) on cooling of the 1.5\%
Cr-doped and pure V\protect\textsubscript{2}O\protect\textsubscript{3}
films with different in-plane lattice parameters engineered by heteroepitaxy.
T\protect\textsubscript{MIT} on cooling is indicated with a star symbol. The grey dashed line in \textbf{a} corresponds to
a 1\% Cr-doped V\protect\textsubscript{2}O\protect\textsubscript{3}
single crystal and in \textbf{b}, to a V\protect\textsubscript{2}O\protect\textsubscript{3}
single crystal (both curves extracted from Ref. \cite{Kuwamoto1980}).
\textbf{c} Room temperaure resistivity (RTR) of the 1.5\% Cr-doped
and pure V\protect\textsubscript{2}O\protect\textsubscript{3} films
as a function of their in-plane lattice parameter.\textbf{ }The blue
and red crosses correspond to the values for a 1\% Cr-doped V\protect\textsubscript{2}O\protect\textsubscript{3}
single crystal in the PM and PI phases, respectively, and the grey
cross to the pure V\protect\textsubscript{2}O\protect\textsubscript{3}
one. Bulk values taken from \cite{Kuwamoto1980} and \cite{Robinson1975}.}
\end{figure*}

The temperature-dependent electrical resistivity curves R(T) of the 1.5\% Cr-doped and pure
V\textsubscript{2}O\textsubscript{3} films are shown in Fig. \ref{Electrical figure}\textbf{a}
and \textbf{b}, respectively, along with the reference curves (grey dashed lines) originating from
1\% Cr-doped and pure V\textsubscript{2}O\textsubscript{3} single crystals \cite{Kuwamoto1980}.
Note that the (Cr\textsubscript{$1-y$}Fe\textsubscript{$y$})\textsubscript{2}O\textsubscript{3}
layers are insulating in agreement with the band gaps of 2.1 eV and 3.2 eV for
(Fe\textsubscript{2}O\textsubscript{3}) and (Cr\textsubscript{2}O\textsubscript{3}) \cite{Chamberlin2013}, respectively,
therefore the transport measurements solely probe the V\textsubscript{2}O\textsubscript{3} thin films.

% Figure Electrical Properties ---------------

%  ---------------

The low temperature (LT) transition can be seen in all samples, except in those with the largest in-plane lattice parameter (dotted red curves in
each figure), which correspond to the highest Fe content $y=1$ in the (Cr\textsubscript{$1-y$}Fe\textsubscript{$y$})\textsubscript{2}O\textsubscript{3} buffer.
The T\textsubscript{MIT} (marked with star symbols in Fig. \ref{Electrical figure}\textbf{a}
and \textbf{b}) was determined from the inflection point in the \emph{dLnR(T)/dT} curve and increased from 150 K to 180 K with the in-plane lattice parameter in both series. This is similar to the change in T\textsubscript{MIT} upon  Cr doping in bulk (see for example the change in T\textsubscript{MIT} in the bulk 1.5\% Cr-doped and pure
V\textsubscript{2}O\textsubscript{3} samples shown in Figs. \ref{Electrical figure}\textbf{a} and \textbf{b}).
Nevertheless, there are differences between the strained films - at constant doping - and bulk samples as a function of doping, such as the lack of sharp transitions and significantly reduced resistance jumps at the transition temperature. These differences can be due to several reasons including the in-plane clamping of the film lattice parameters that hampers the structural phase transitions as well as small variations in composition such as Cr-doping and oxygen content. It has been previously reported that increase of tensile strain \cite{Dillemans2012} or oxygen deficiency \cite{Brockman2011} can cause an increase of RTR and a decrease of resistance ratio across the MIT. Therefore, possible formation of oxygen vacancies  could explain the crossing of the R(T) curve for the most strained films (red dotted curves) with the one of films with smaller lattice parameter (yellow dotted curves).

The PM-PI transition close to RT is, as expected, completely absent in the 1.5\% Cr-doped V\textsubscript{2}O\textsubscript{3}
films for all in-plane parameters, most probably due to the clamping effect \cite{Homm2015}.
Nonetheless, a significant change of the RTR occurs when changing the in-plane lattice parameter of the films.
This trend is also observed in the pure V\textsubscript{2}O\textsubscript{3}
thin film series (Fig. \ref{Electrical figure}\textbf{b}).
Figure \ref{Electrical figure}\textbf{c} highlights the RTR values as a function of the in-plane lattice parameter for both film series, which clearly shows a PM to PI transition triggered by epitaxial strain.
The most striking result is that an in-plane lattice deformation of \textasciitilde{} 1\%  can trigger a $\Delta$R/R
as large as 100,000 \% in both the strained 1.5\% Cr-doped (yellow circles) and the pure (green circles) V\textsubscript{2}O\textsubscript{3} films.
To our knowledge this is the largest change in RTR achieved by epitaxial strain in  V\textsubscript{2}O\textsubscript{3} compounds \cite{Salev2019,Alyabyeva2018,Sakai2019}.

The RTR of the 1.5\% Cr-doped V\textsubscript{2}O\textsubscript{3}
film with the smallest in-plane lattice parameter (\textit{a} = 4.957 \AA) is closer to the bulk RTR value of the PM phase of pure V\textsubscript{2}O\textsubscript{3}
(marked by the grey cross PM\textsuperscript{undoped}) than to the one of the PM phase of 1\% Cr-doped V\textsubscript{2}O\textsubscript{3}
compound (marked by the blue cross PM\textsuperscript{1\% Cr}).
This low RTR value suggests that the metallic state achieved by strain in 1.5\% Cr-doped
V\textsubscript{2}O\textsubscript{3} thin films is mostly determined by a collective modification of the crystal lattice and electronic band structure in contrast to bulk samples, where the non-perfectly homogeneous distribution of Cr dopants can lead to electron scattering and to higher RTR.

%\subsection{Electrodynamics of the strained V\textsubscript{2}O\textsubscript{3} thin film compounds}

\begin{figure*}
\includegraphics[scale=0.35]{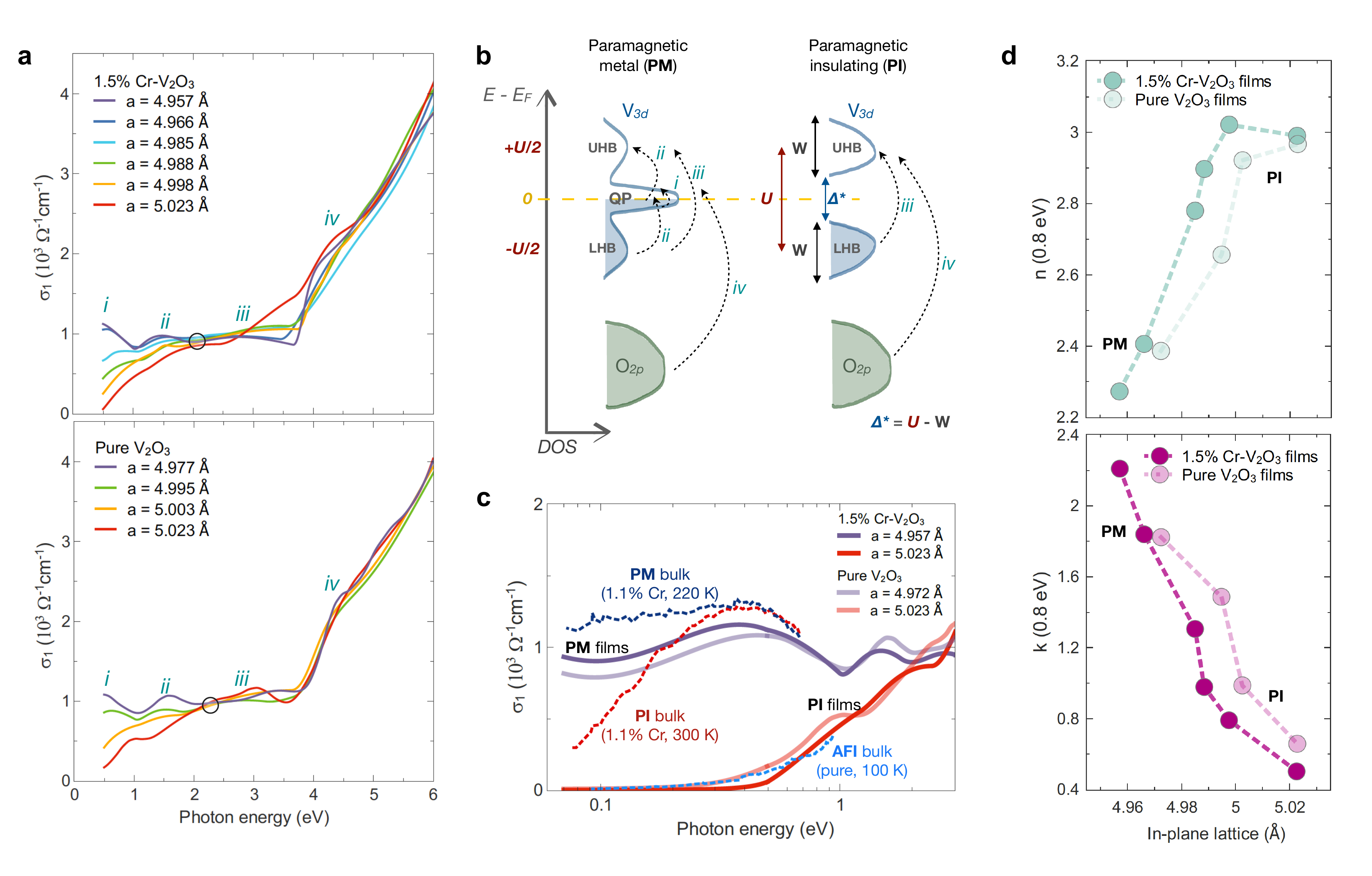}
\caption{\label{Optical figure}Spectral optical properties of strained 1.5\%
Cr-doped and pure V\protect\textsubscript{2}O\protect\textsubscript{3}
thin films. \textbf{a} Real part of the optical conductivity ($\sigma$\protect\textsubscript{1})
as a function of photon energy of the strained 1.5\% Cr-doped (top)
and pure V\protect\textsubscript{2}O\protect\textsubscript{3} (bottom)
thin film series. Interband transitions are labelled \textit{i} -
\textit{iv}, according to schematics in \textbf{b}. The circles indicate
isosbestic points at around 2.1 eV in both series. \textbf{b} Sketch
of the electronic density of states (DOS) and optical excitations
in the PM and PI phases of V\protect\textsubscript{2}O\protect\textsubscript{3}
compounds based on the model for the optical conductivity in Mott-Hubbard
systems \cite{Rozenberg1995}. \textbf{c} Optical conductivity curves
extended down to 70 meV in the mid-infrared (MIR) for the films with
the extreme in-plane lattice parameters of each series. The dashed
dark blue and red curves correspond to the PM and PI spectra of a
1.1\% Cr-V\protect\textsubscript{2}O\protect\textsubscript{3} single
crystal, respectively, and the light blue one to the AFI spectrum
of a pure V\protect\textsubscript{2}O\protect\textsubscript{3} single
crystal. Bulk data taken from Ref. \cite{Lupi2010,Vecchio2015}.
\textbf{d} Optical constants n (top) and k\textit{ }(bottom) at the
photon energy of 0.8 eV ($\lambda$= 1550 nm) of each series as
a function of the engineered in-plane lattice parameter. }
\end{figure*}

Optical studies have been invaluable for understanding the underlying physics in V\textsubscript{2}O\textsubscript{3}
systems \cite{Lupi2010,Vecchio2015,Barker1970,Stewart2012,Baldassarre2008,Qazilbash2007} because electron correlation effects appear as distinct signatures in the optical conductivity spectra of metallic
and insulating phases and the Mott MIT is accompanied by large spectral weight changes \cite{Rozenberg1995,Brahlek2017}.
The optical conductivity spectra are shown in Fig. \ref{Optical figure}\textbf{a} for the strained
1.5\% Cr-doped (top) and pure (bottom) V\textsubscript{2}O\textsubscript{3} films.
The features (labelled from \textit{i} to \textit{iv}) are assigned to the optical transitions depicted in
Fig. \ref{Optical figure}\textbf{b}.
The scheme in Fig. \ref{Optical figure}\textbf{b}
shows the density of states (DOS) of metallic (PM) and insulating
(PI) phases based on the one-band Hubbard model \cite{Rozenberg1995},
in which the vanadium 3\textit{d} band is composed of the quasiparticle
peak (QP) around the Fermi level (in the PM phase, defining feature
\textit{i}) and the incoherent bands (so\textendash called Low Hubbard Band (LHB) and Upper Hubbard Band
(UHB)) emerging at \textit{E - E}\textsubscript{\textit{F}}\textit{ = \textpm{} U/2}, each of
width W and separated by the Coulomb energy \textit{U}.
Note that the energy scale of the optical transitions (\textit{ii}) and (\textit{iii})
are in the order of \textit{U/2} and \textit{U}, respectively, and
the Mott gap in the insulating state is in the order of $\Delta${*}
= \textit{U} - W.

Although in the case of V\textsubscript{2}O\textsubscript{3} the realistic multi-orbital scenario of the V \textit{3d}
bands (\textit{i.e.}, Hubbard bands and QP are composed of two different
orbitals) makes the DOS and optical transitions
much more complex \cite{Stewart2012,Tomczak2009}, the optical responses
across the Mott MIT can be explained semiquantitatively based on the
one-band Hubbard model for the optical conductivity \cite{Rozenberg1995}. 
The real part of the optical conductivity $\sigma$\textsubscript{1},
defined as $\sigma$\textsubscript{1}{\small{}($\omega$)
= 2$\epsilon$}\textsubscript{{\footnotesize{}0}}nk,{\small{}
}with {\small{}$\omega$} the angular frequency and {\small{}$\epsilon$}\textsubscript{{\footnotesize{}0}}
the vacuum permittivity \cite{Opticalbook}, has been calculated
from the optical constants (index of refraction\textit{, }n and extinction
coefficient, k) derived by Spectroscopic Ellipsometry (SE) experiments
performed at RT in an energy range from 0.5 eV to 6 eV (see supplementary information for details).

Consistent with the one-band Hubbard model \cite{Rozenberg1995}, a signature of the Drude conductivity
(feature \textit{i}, associated with optical transitions within the
quasiparticle (QP) peak in the PM phase) is spotted at low energies
in both film series with the smallest in-plane lattice parameters. The effects of electron
correlation appear in the visible range as features \textit{ii} (\textasciitilde{}
1.4 eV in film spectra) and \textit{iii} (\textasciitilde{} 2.8 eV
in film spectra), associated to optical transitions between the QP
peak and the Hubbard bands, and between LHB and UHB, respectively.
Finally, the peak around 4 eV (labelled as \textit{iv}) observed in
all films is related to transitions from the oxygen 2p bands.

As can be seen in Fig. \ref{Optical figure}\textbf{a}, the features below 2 eV (\textit{i} and \textit{ii}) in $\sigma$\textsubscript{1} significantly decrease, when moving from the PM to the PI state triggered by epitaxial strain,
owing to the spectral weight transferred between the Hubbard bands and the QP.
The drastic decrease in $\sigma$\textsubscript{1}
in this low energy range for the insulating films is directly related
to the suppression of the QP and suggests the gradual opening of a
gap across the Mott MIT in both strained V\textsubscript{2}O\textsubscript{3} thin film series.

The data in Fig. \ref{Optical figure}\textbf{a} also reveal isosbestic points near 2.1 eV in both series (marked by circles).
Such a point is a characteristic feature of Mott transitions \cite{Eckstein2007},
which has been observed in V\textsubscript{2}O\textsubscript{3} thin films, single crystals \cite{Stewart2012,Baldassarre2008}
and VO\textsubscript{2} films \cite{Qazilbash2007,Li2011} but for temperature-dependent optical conductivity spectra, which intersect at the same $\sigma$\textsubscript{1}{\small{} }and energy values due to
redistributions of spectral weight associated with a change from itinerant to localized electrons.
Consequently, the presence of an isosbestic point in the RT optical conductivity spectra of both strained film series further suggests that a Mott phase transition can be triggered by continuous lattice deformation induced by heteroepitaxy.

In Fig. \ref{Optical figure}\textbf{c} the optical conductivity spectra extended down to  the mid-infrared (MIR) regime of the 1.5\% Cr-doped
and pure V\textsubscript{2}O\textsubscript{3} films with the smallest and largest in-plane lattice parameters are shown.
For comparison, the optical conductivity curves obtained from single crystals (1.1\% Cr-doped V\textsubscript{2}O\textsubscript{3} single crystal in the PM (220 K) and PI (300 K) phases and pure V\textsubscript{2}O\textsubscript{3}
single crystal in the AFI (100 K) phase) \cite{Lupi2010,Vecchio2015} are added.
These bulk data illustrate that there is only a small drop in spectral weight below 1 eV linked to the opening of a band gap
of \ensuremath{\sim} 0.12 - 0.2 eV \cite{Mo2006} for the Mott PM-PI transition in bulk.
In contrast, the changes in the strained films are spectacular with a significant spectral
weight drop and opening of much larger band gap of up to \ensuremath{\sim} 0.5 eV.
There has been no comparable precedent report in the literature. The gradual decrease of $\sigma$\textsubscript{1} for energies below 1 eV as the lattice parameter increases, and the films become more insulating, indicates that the large spectral weight at low energies  might not be an intrinsic property of the PI phase. In addition, it is remarkable
that the evolution of $\sigma$\textsubscript{1} of the more insulating films resembles that of the reference V\textsubscript{2}O\textsubscript{3} single crystals in the AFI phase, for which a large band gap of 0.6 eV opens \cite{Thomas1994}. Our optical data (features \textit{ii} and \textit{iii}) indicate
that \textit{U} in the films is of the same order as in bulk. Therefore, a bandwidth reduction as a function of \textit{a}-axis expansion is possibly the origin of the larger gap in the PI phase  in films with \textit{a} > 4.985 \AA , as the Mott gap is $\Delta${*} = \textit{U} - W.

% Figure Optical Properties ---------------

% ---------------

The modulation of the bandgap, associated with intermediate states
stabilized across the Mott transition, allows a considerable range
of tuning of the optical constants at RT.
For example, Fig. \ref{Optical figure}\textbf{d}
displays the optical constants n (top) and k\textit{ }(bottom) at
the photon energy of 0.8 eV \textendash which corresponds to a wavelength
of 1550 nm typically used in telecommunications\textendash{} across the Mott
transition at RT in both series.
At this photon energy, n varies between
\ensuremath{\sim} 2.3 and \ensuremath{\sim} 3 while k changes between
\ensuremath{\sim} 2.2 and \ensuremath{\sim} 0.5 as a function of the in-plane lattice parameter. These trends and
values are similar to those reported for the temperature driven MIT in VO\textsubscript{2} \cite{Li2011,Kana2011}, however, the transition in the latter occurs at about 70 $^{\circ}$C whereas our films show a transition at RT, which is beneficial for device applications.
It also has to be noted that the optical constants for V\textsubscript{2}O\textsubscript{3}
compounds have not been reported previously. For example, the n and k values of
the 1.5\% Cr-doped V\textsubscript{2}O\textsubscript{3}
films in the whole NIR-UV energy range can be found in Fig. S2.4 of the supplementary information.

% Figure Phase Diagram ---------------

\begin{figure*}[h]
\includegraphics[scale=0.35]{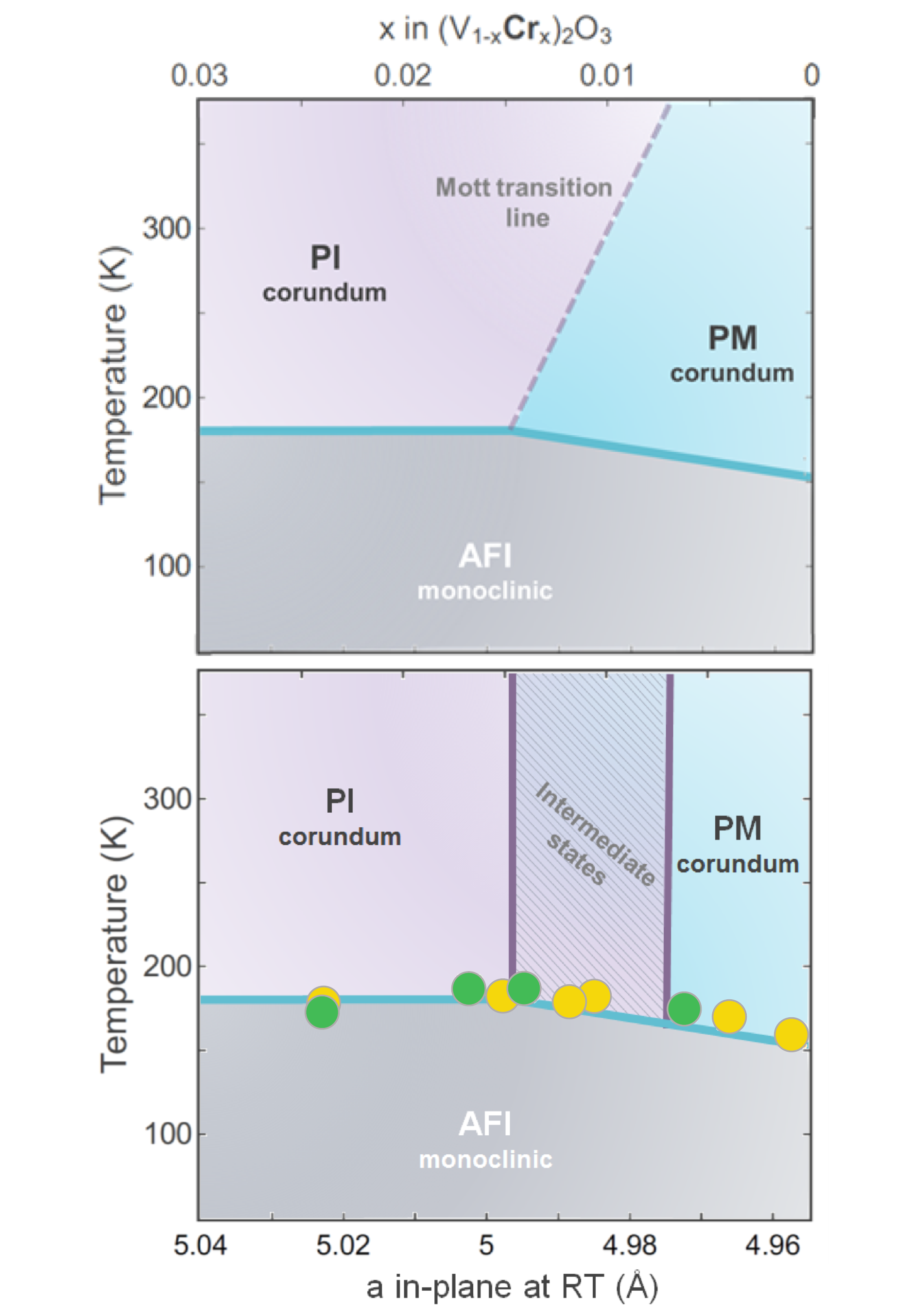}
\caption{\label{Phase diagram}(top) \textit{Temperature - Cr-doping} phase diagram of bulk V\protect\textsubscript{2}O\protect\textsubscript{3} (adapted from \cite{McWhan1973}) The dotted line corresponds
to the Mott PM-PI transition  and the continuous line the LT transition to the AFI phase. (bottom) Proposed \textit{Temperature - In-plane lattice parameter} (a in-plane at RT) phase diagram for pure and Cr-doped  V\protect\textsubscript{2}O\protect\textsubscript{3} thin films. A pure Mott PM-PI phase transition at high temperatures (200K - 350K) is indicated by the vertical continuous line. 
The shaded area indicates the intermediate phases accessible by strain engineering the in-plane lattice parameter. The green and yellow
circles correspond to the low temperature transitions to the AFI phase in pure and 1.5\% Cr-V\protect\textsubscript{2}O\protect\textsubscript{3} thin films, respectively. }
\end{figure*}

%  ---------------

The structural, electrical and optical results reported here for pure and Cr-doped V\textsubscript{2}O\textsubscript{3} thin films indicate that the well-known temperature - doping phase diagram (see Fig. \ref{Phase diagram}(top)) for bulk V\textsubscript{2}O\textsubscript{3} can not be used to represent the different phases achieved via strain engineering. 
Figure \ref{Phase diagram}(bottom) shows the proposed phase diagram where temperature and in-plane lattice parameter are the variable axes.
The absence of a PM-PI phase transition with temperature  in pure and 1.5\% Cr-doped V\textsubscript{2}O\textsubscript{3} thin films suggests that the 1st order Mott PM-PI transition line (dashed line in Fig. \ref{Phase diagram}(top)) in the  bulk phase diagram should be replaced by two straight lines comprising a region of intermediate states between the stable PM and PI ones. These intermediate states are characterized by lattice parameter, RTR and optical constant values between the bulk PM and PI ones. Although these results point to the stabilization of homogeneous intermediate states, local techniques will be necessary to investigate whether there is phase coexistence at nanometric scales.These novel intermediate phases, accessible thanks to the systematic control of strain by heteroepitaxy, can be very interesting for applications based on V\textsubscript{2}O\textsubscript{3} compounds. For example, the  changes in resistivity vs lattice parameter could be exploited by inducing lattice deformations when applying an electric field to a nearby piezoelectric material \cite{Salev2019} or by depositing the material on a membrane, the effect can be used as a piezoresistive transduction mechanism.

The observation of a PM-PI phase transition by strain engineering in a wide range of temperatures (around room temperature) shows that the Mott transition can be solely driven by lattice deformation. In contrast, the temperature-driven transition to the AFI phase (yellow and green circles in Fig. \ref{Phase diagram}(bottom)) is preserved for the whole range of  in-plane lattice parameters studied in this work \cite{TMIT}. Notably, the temperature dependence of the boundary between the PM (or PI) and AFI  phases resembles the one for the bulk samples.
The robustness of both PM-AFI and PI-AFI transitions against  substrate clamping can be related to the considerably larger thermodynamic enthalpy of formation (\emph{$\Delta$H} \emph{=} 2.03 kJ/mole and 2.36 kJ/mole, respectively) as compared to the enthalpy of formation for the PM-PI (\emph{$\Delta$H} \emph{=} 0.15 kJ/mole) transition \cite{Keer1976,Keer1977}.

\section{Conclusion}

In this work, the direct connection between the structure, electronic
and optical properties has been exploited to control the Mott MIT
phase transition at room temperature in V\textsubscript{2}O\textsubscript{3} thin film compounds.
Intermediate states between PM and PI phases, which are
not accessible in bulk, are realized  by "smart"
heteroepitaxy based on fabrication of tailored artificial substrates. It is
demonstrated that the Mott MIT can be induced at RT both, in pure and Cr-doped compounds, by strain
engineering, leading to films with a colossal $\Delta$R/R
$\sim$100,000 \% and a broad range of optical constants.
These results demonstrate that  Cr doping is actually not the key control parameter for the RT Mott MIT, in contrast to the observations in the bulk.
Moreover, strain engineering leads to a more homogeneous modification of Mott materials compared to chemical doping, where the modification occurs locally on specific sites.

The observed unique electrical and optical properties are attributed to the strain-modulated bandgap beyond the bulk value of 0.1 eV.
Our data suggest that the
significant spectral weight at low energies in the bulk PI phase,
traditionally considered as a hallmark of Mott physics \cite{Hansmann2013,Guo2014,Basov2011,Kotliar1999},
might not be an intrinsic feature but an artefact owing to local inhomogeneities.
X-ray absorption  and Photo-emission spectroscopy studies together with theoretical calculations considering
the lattice deformations and local structures, could give more detailed insight into the changes in the electronic bandstructure and the origin of the modulated bandgap in the strained films. 

Finally, the stabilization of a Mott PM-PI transition, in both pure and Cr-doped V\textsubscript{2}O\textsubscript{3}  thin films, around room temperature and the possibility to modify the structural, electrical and optical properties  in a continuous manner  can be used in more practical concepts such as pressure sensors, resonators, acoustic light modulators, and photonic devices in the near-infrared range.

\section*{Author contributions }

P.H. performed all the MBE thin film growth experiments and the X-ray
diffraction measurements and analysis. P.H. and M.M. performed the
electrical measurements and analysis. J.W.S did the TEM and EDX analysis.
S.P. measured the SE spectra and fitted the data to extract the optical
constants. P.H. performed the SE data analysis. J.P.L. conceived the
underlying concept and experiments. J.P.L, P.H., M.M. and J. W. S. interpreted
the data and wrote the final version. All authors contributed to the
writing of the manuscript and the scientific understanding.

\section*{Acknowledgments}

The authors acknowledge financial support from the EU H2020 Project 688579 PHRESCO and the KU Leuven projects GOA/13/011
Fundamental challenges in Semiconductors and C14/17/080 2D Oxides. We thank Dr. Ruben Lieten for the program to analyze reciprocal space maps data and Dr. Ilse Lenaerts for critically reading the manuscript. P.H. acknowledges support from Becas Chile - CONICYT.

\section*{Data availability}
The data that support the findings of this study are available from the corresponding authors upon reasonable request.

\section*{Supplementary information}

The Supporting Information contains X-ray diffraction data of the Cr-doped  V\textsubscript{2}O\textsubscript{3} series, an example of the steps followed to fit the Spectroscopic Ellipsometry data and the resulting optical constants of  the Cr-doped  V\textsubscript{2}O\textsubscript{3} series in the energy range from 0.5 to 6 eV.


\begin{thebibliography}{9}
\bibitem{TMIT}{The transition temperature for the films with the largest in-plane lattice parameters were obtained from optical transmission measurements.}
\end{thebibliography}
\end{document}